\documentclass[smallextended]{svjour3} 
\smartqed 
\usepackage{graphicx,bm}
\usepackage{dsfont} 
\newcommand{\be}{\begin{equation}}
\newcommand{\ee}{\end{equation}}
\newcommand{\bea}{\begin{eqnarray}}
\newcommand{\eea}{\end{eqnarray}} 
\newcommand{\nab}{\nabla}
\newcommand{\nn}{\nonumber}
\newcommand{\rf}[1]{(\ref{#1})}
\newcommand{\bra}[1]{\left(#1\right)}
\newcommand{\bras}[1]{\left[#1\right]}
\newcommand{\brac}[1]{\left\{#1\right\}}
\newcommand \veps {\varepsilon}
\newcommand{\lb}{\{}
\newcommand{\rb}{\}}
\newcommand{\A}{{\cal A}}
\newcommand{\E}{{\cal E}}
\renewcommand{\H}{{\cal H}}
\newlength{\dhatheight}
\newcommand{\doublehat}[1]{%
    \settoheight{\dhatheight}{\ensuremath{\hat{#1}}}%
    \addtolength{\dhatheight}{-0.35ex}%
    \hat{\vphantom{\rule{1pt}{\dhatheight}}%
    \smash{\hat{#1}}}}
\begin{document}
\title{The 1+1+2 formalism for  Scalar-Tensor gravity}
\author{Sante Carloni \and Peter K S Dunsby }
\institute{Sante Carloni \at Centro Multidisciplinar de Astrofisica - CENTRA,
Instituto Superior Tecnico - IST,
Universidade de Lisboa - UL,
Avenida Rovisco Pais 1, 1049-001, Portugal \and Peter K S Dunsby \at Department of Mathematics and Applied\ Mathematics,
University of Cape Town, South Africa, \and South African
Astronomical Observatory, Observatory Cape Town, South Africa }
\maketitle
\begin{abstract}
We use the 1+1+2 covariant approach to clarify a number of aspects of spherically symmetric solutions of non-minimally coupled scalar tensor theories. Particular attention is focused on the extension of Birkhoff's theorem and the nature of quasi-local horizons in this context.
\end{abstract}

\tolerance=5000

\section{Introduction}
The term {\em scalar-tensor} (ST) theory can be found in literature to describe theories of gravity in which either minimally or  non-minimally coupled scalar fields appear. The first class of theories are the most studied,  due in part to their 
simplicity and connection to the original models of inflation. The second class, for which the term ST is mostly used, has instead gained attention when it was discovered they arise 
naturally in the context of quantum field theory on curved spacetime \cite{Donoghue} and of higher dimensional theories such as Kaluza-Klein and (Super-)String theories \cite{Bergmann}. This type of ST theories is currently one the most studied classes of extensions of General Relativity (GR). For minimally and non-minimally coupled theories, we have now a relatively clear picture of many aspects of the cosmologies of these models \cite{Fuji-Maeda,FaraoniBook}, including applications to inflation \cite{LaStein}  and to the modeling the late-time acceleration of the Universe  \cite{ScTnDarkEnergy}. In relativistic astrophysics, ST theories are considered for example in gravitational wave physics and in the analysis of deviations from GR in the context of neutron stars \cite{Damour:1993hw}. These theories have also played a role in the debate about the no-hair conjecture of black holes and its various realizations \cite{Bekenstein} and have been discovered to give rise some peculiar phenomenology in the GR limit \cite{GRLimit}.

In spite of these results our knowledge of some fundamental aspects of ST theories such as the existence and nature of exact static and spherically symmetric solutions is still limited. This is particularly true for theories which involve a non-trivial self-interacting potential. Indeed, many of the most important models of ST theories are characterized by a non-trivial self-interaction of the scalar field potential and, especially in the  non-minimally coupled case,  very few solutions in the Jordan frame are known \cite{Brans}. 

In the simpler context of cosmology a number of formalisms have been used to overcome similar problems in the context of GR. In particular the 1+3 covariant formalism originally developed by Ehlers and Ellis \cite{Covariant}  has been (and it is still now) particularly successful in the analysis of complex  anisotropic and/or non homogenous cosmological models \cite{ellisbook}. In more recent times, the same formalism has been generalised to deal with the cosmology of several extensions of general relativity, including ST theories. Its application has led to a number of important results in the understanding of the dynamical of homogeneous cosmological models as well as the dynamics of scalar perturbations for ST theories \cite{SanteST}. The 1+3 covariant approach can be also extended to treat spacetimes which are less symmetric than cosmological ones. For example the 1+1+2 formalism, developed by Clarkson and Barrett \cite{extension}, was applied in the context of GR to the general study of LRS spacetimes (see below for more precise definition of these spacetimes), the analysis of the linear perturbations of the Schwarzschild spacetime \cite{extension}, gravitational lensing \cite{LensingST} and to the generation of electromagnetic radiation by gravitational waves interacting with a strong magnetic field around a vibrating Schwarzschild black hole \cite{Betschart}. 

In this paper we generalize this approach to the case of ST theories with the aim to obtain a deeper insight into the properties of spherically symmetric solutions of these theories. The class of non-minimally coupled theories we will consider are those characterised by a standard kinetic term for the scalar field. However, the actions we treat can be recast in the general form given by Bergmann, Nordtveldt and others \cite{Nordtvedt} by a simple reparameterisation of the scalar field. 

As we will we see the application of the 1+1+2 formalism offers an alternative point of view that can help to clarify some known issues of the physics of ST theories. Most importantly, the 1+1+2 formulation of the field equations of ST theories represents the starting point to their systematic investigation in the context of LRS-II spacetimes and the results obtained can be crucial for the comparison of these theories with the observations. 

The paper is organised  in the following way. In section 2 we give the general equations of ST gravity. In Section 3 we briefly review the covariant approach and deduce the 1+1+2 framework from the 1+3 one. We then derive the 1+1+2 equations in general and relate the 1+1+2 quantities to the metric components. In section 4 we give the 1+1+2 equations for a general non-minimally coupled ST theory of gravity and we present some important results which are needed if one is to use 1+1+2 formalism to find exact solutions. In section 5 we discuss  a number of aspects of ST theories in the framework of the 1+1+2 approach like the existence and the meaning of Birkhoff's  theorem and the existence of quasi-local horizons. Finally section 6 is dedicated to our conclusions.

Unless otherwise specified, natural units ($\hbar=c=k_{B}=8\pi G=1$) will be used throughout this paper and Latin indices run from 0 to 3. The symbol $\nabla$ represents the usual covariant derivative and $\partial$ corresponds to partial differentiation. We use the $-,+,+,+$ signature and the Riemann tensor is defined by
\begin{equation}
R^{a}{}_{bcd}=\Gamma^a{}_{bd,c}-\Gamma^a{}_{bc,d}+ \Gamma^e{}_{bd}\Gamma^a{}_{ce}-\Gamma^e{}_{bc}\Gamma^a{}_{de}\;,
\end{equation}
where the $\Gamma^a{}_{bd}$ are the Christoffel symbols (i.e. symmetric in the lower indices), defined by
\begin{equation}
\Gamma^a_{bd}=\frac{1}{2}g^{ae}
\left(g_{be,d}+g_{ed,b}-g_{bd,e}\right)\;.
\end{equation}
The Ricci tensor is obtained by contracting the {\em first} and the {\em third} indices
\begin{equation}\label{Ricci}
R_{ab}=g^{cd}R_{acbd}\;.
\end{equation}
Symmetrisation and the anti-symmetrisation over the indexes of a tensor are defined as 
\begin{equation}
T_{(a b)}= \frac{1}{2}\left(T_{a b}+T_{b a}\right)\;,~~ T_{[a b]}= \frac{1}{2}\left(T_{a b}-T_{b a}\right)\,.
\end{equation}
Finally the Hilbert--Einstein action in the presence of matter is given by
\begin{equation}
{\cal A}=\frac12\int d^4x \sqrt{-g}\left[R+ 2{\cal L}_m \right]\;.
\end{equation}
\section{General equations for Scalar Tensor Gravity} \label{GenST}
The  action for ST theories of gravity is given by (conventions as in Wald \cite{bi:wald})\footnote{ For the sake of simplicity, we do not consider here more complex generalisations of ST, which include Galileon fields \cite{Galileon} or  Horndeski theories \cite{Horndeski}. The extension of our approach to these contexts is, however, straightforward.}:
\begin{eqnarray}\label{eq:actionScTn}
&&\mathcal{A}=\int d x^{4}\sqrt{-g}\left[\frac{1}{2}F(\psi)R-\mathcal{L}_\psi+\mathcal{L}_m \right],
\end{eqnarray}
where 
\begin{eqnarray}
&&\mathcal{L}_\psi=\frac{1}{2}\nab_a\psi\nab^a\psi +V(\psi),
\end{eqnarray}
$V(\psi)$ is  a generic potential expressing the self-interaction of the scalar field and $\mathcal{L}_m$ 
represents the matter contribution.

Varying the action with respect to the metric gives the gravitational
field equations:
\begin{eqnarray}
  F(\psi)\left(R_{ab}-\frac{1}{2}Rg_{ab}\right)&=&T _{ab}^{m}+\nab_a\psi\nab_b\psi -g_{ab}
\left(\frac{1}{2}\nab_c\psi\nab^c\psi+V(\psi)\right)+\nn\\ &&\left(\nab_b\nab_a - g_{ab}\nab_c\nab^c\right)F(\psi)\;,
\label{eq:einstScTn}
\end{eqnarray}
and the variation with respect to the field $\psi$ gives the
curved spacetime version of the Klein--Gordon equation
\begin{equation}
\nab_a\nab^a\psi+\frac{1}{2}F'(\psi) R -V'(\psi)=0\;, \label{eq:KG}
\end{equation}
where the prime indicates a derivative with respect to $\psi$.
Both these equations reduce to the standard equations for GR and a
minimally-coupled scalar field when $F(\psi)=1$.

Equation (\ref{eq:einstScTn}) can be recast as
\begin{equation}
\label{eq:einstScTneff}
 G_{ab}=\frac{ T_{ab}^{m}}{F(\psi)}+T^{\psi}_{ab}=T^{(eff)}_{ab}\,,
 \end{equation}
where $T^{\psi}_{ab}$ has the form
\begin{eqnarray}\label{eq:TenergymomentuEff}
T_{ab}^{\psi}&=&\frac{1}{F(\psi)}\left[\nab_a\psi\nab_b\psi-g_{ab}
\left(\frac{1}{2}\nab_c\psi\nab^c\psi+V(\psi)\right)\right.\nn\\&&\left.
 +\nab_b\nab_aF(\psi)- g_{ab}\nab_c\nab^cF(\psi)\right]\;. \label{eq:semt}
\end{eqnarray}
Provided  that $\psi_{,a} \neq 0$, equation (\ref{eq:KG}) also
follows from the conservation equations
\begin{equation}
\nab^bT_{ab}^{eff}=0\;.
 \label{eq:cons}
\end{equation}
The reformulation above will be very important  for our purposes. In fact, the form of
(\ref{eq:einstScTneff}) allows us to treat scalar tensor gravity
as standard Einstein gravity in presence of two effective fluids and permits a straightforward generalisation of the 1+1+2 formalism to these equations.
\section{ The 1+1+2 Covariant Approach}
In the following we give a very brief review of the 1+1+2 covariant approach (for details see \cite{extension}). We will proceed first with the standard 1+3 decomposition and then perform a further split of the spatial degrees of freedom relative to a preferred spatial direction. This allows us to derive a set of variables better suited to systems in which a spatial direction is important (i.e., the radial one in the case of spherical symmetry).
\subsection{Kinematics} 
In 1+3 approach we define a time-like congruence, with unit tangent vector $u^a$ ($u^au_a=-1$).  In this way, any tensor field can be projected along $u^a$ (extracting the temporal parts) or into the 3-space orthogonal to $u^a$ using the projection tensor $h^a_b=g^a_b+u^au_b$.

In the 1+1+2 approach, we further split this 3-space by introducing the spatial unit vector $e^{a}$ orthogonal to $ u^{a}$, so that
\be
e_{a} u^{a} = 0\;,\; \quad e_{a} e^{a} = 1\;.
\ee
Then the tensor
\be 
N_{a}{}^{b} \equiv h_{a}{}^{b} - e_{a}e^{b} = g_{a}{}^{b} + u_{a}u^{b} 
- e_{a}e^{b}~,~~N^{a}{}_{a} = 2 
\label{projT} 
\ee 
projects vectors into the 2-surfaces orthogonal to $e^{a}$ \textit{and} $u^a$. It is obvious that $e^aN_{ab} = 0 =u^{a}N_{ab}$. Using $N_{ab}$, any 3-vector $\lambda^{a}=h^a_b\lambda^b$ can be irreducibly split into a component along $e^{a}$ and a sheet component $\Lambda^{a}$, orthogonal to $e^{a}$, i.e., 
\be
\lambda^{a} = \Lambda e^{a} + \Lambda^{a}\,, \quad \Lambda \equiv\lambda^{a} 
e_{a}\,, \quad\Lambda^{a} \equiv N^{ab}\lambda_{b}\;.
\label{equation1} 
\ee 
A similar decomposition can be done for symmetric trace free 3-tensors $\lambda_{ab}=h^c_ah^d_b\Phi_{cd}$, which can be split into scalar (along $e^a$), 2-vector and 2-tensor parts as follows:
\be 
\lambda_{ab} = \lambda_{\langle ab\rangle } = \Lambda\bra{e_{a}e_{b} - \frac{1}{2}N_{ab}}+
2 \Lambda_{(a}e_{b)} + \Lambda_{ab}\;, 
\label{equation2} 
\ee 
where 
\bea
\Lambda &\equiv & e^{a}e^{b}\lambda_{ab} = -N^{ab}\lambda_{ab}\;,\nonumber \\ 
\Lambda_{a} &\equiv & N_{a}{}^be^c\lambda_{bc}\;,\nonumber \\ 
\Lambda_{ab} &\equiv & \lambda_{\brac{{ab}}} 
\equiv \bra{ N^{c}{}_{(a}N_{b)}{}^{d} - \frac{1}{2}N_{ab} N^{cd}} \lambda_{cd}\;.
\eea 
The  Levi-Civita 2-tensor plays the usual role of the 2 volume element for the 2-surfaces and it is defined as
\be
\veps_{ab}\equiv\veps_{abc}e^{c} = \eta_{dabc}e^{c}u^{d}\;, \label{perm}
\ee 
where $\veps_{abc}$ is the 3-space permutation tensor, which is the volume element of the 3-space and $\eta_{abcd}$ is  the spacetime 4-volume element.

With these definitions it follows that any 1+3 quantity can be locally split into three types of objects: scalars, 2-vectors and  2-tensors defined on the 2-surfaces orthogonal to $e^a$. 
\subsection{Derivatives and the kinematical variables} 
Using $u^a$ and $h_{ab}$ we can obtain two derivative operators: one defined along the time-like congruence:
\be
\dot{X}^{a..b}{}_{c..d}{} = u^{e} \nab_{e} {X}^{a..b}{}_{c..d}
\ee
and the projected  derivative $D$: 
\be
D_{e}X^{a..b}{}_{c..d}{} = h^a{}_f h^p{}_c...h^b{}_g h^q{}_d h^r{}_e \nab_{r} {T}^{f..g}{}_{p..q}\;.
\ee
Applying the covariant derivative to $u_a$ we can obtain the key 1+3 quantities:
\be 
\nabla_{a}u_{b} = -u_{a}\dot{u}_{b} + \frac{1}{3}\Theta h_{ab} + 
\sigma_{ab} + \omega_{ab}\;,
\ee
where $\dot{u}_a$ is the acceleration, $\Theta$ is the expansion parameter, $\sigma_{ab}$ the
shear and $\omega_{ab}$ is the vorticity.

In the same way as before we can now split the $D$ operator using $e^{a}$ and $N_{ab}$:
\bea
\hat{X}_{a..b}{}^{c..d} &\equiv & e^{f}D_{f}X_{a..b}{}^{c..d}\;, 
\\
\delta_f X_{a..b}{}^{c..d} &\equiv & N_{a}{}^{e}..N_{b}{}^gN_{i}{}^{c}..
N_{k}{}^{d}N^j{}_fD_j X_{e..g}{}^{i..k}\;.
\eea 
The covariant derivative of $e^a$ can be split in the direction orthogonal to $u^a$ into it's irreducible parts to give: 
\be 
{\rm D}_{a}e_{b} = e_{a}\alpha_{b} + \frac{1}{2}\phi N_{ab} + 
\xi\veps_{ab} + \lambda_{ab}\;.
\ee
For an observer that chooses $e^{a}$ as a special direction in spacetime, $\phi=\delta_a e^a$ represents the \textit{expansion of the sheet},  $\lambda_{ab}= \delta_{\lb a}e_{b \rb }$ is the \textit{shear of $e^{a}$} (i.e., the distortion of the sheet), $\xi=\frac{1}{2} \veps^{ab}\delta_{a}e_{b}$  is a representation of the ``twisting'' or rotation of the sheet and $\alpha_{a}= \hat{e}_{a}$ its \textit{acceleration}.
 
Using equations (\ref{equation1})  and (\ref{equation2}) one can split the 1+3 kinematical variables and Weyl tensors as follows:
\bea 
\Theta&=& -\alpha_bu^b+\delta_a u^a\\
\dot{u}^{a} &=& \A e^{a}+ \A^{a}\;,\label{1+1+2Acc} \\ 
\omega^{a}&=& \frac{1}{2}\veps^{abc}\omega_{bc} = \Omega e^{a} +\Omega^{a}\;,\\ 
\sigma_{ab} &=& \Sigma\bra{ e_ae_b - \frac{1}{2}N_{ab}} + 
2\Sigma_{(a}e_{b)} + \Sigma_{ab}\;, \\ 
E_{ab}&=&C_{abcd} u^c u^d= \E\bra{ e_{a}e_{b} - \frac{1}{2}N_{ab}} + 
2\E_{(a}e_{b)} + \E_{ab}\;, \\
H_{ab}&=&\frac{1}{2}\varepsilon_{ade}C^{de}{}_{bc} u^c = \H\bra{ e_{a}e_{b} - \frac{1}{2}N_{ab}} + 2\H_{(a}e_{b)} + 
\H_{ab}\;,\label{MagH} 
\eea 
where $E_{ab}$ and $H_{ab}$ are the electric and magnetic part of the 
Weyl tensor respectively. It follows that the key variables of the 1+1+2 formalism are:
\be
\brac{ \Theta, \A, \Omega,\Sigma, \E, \H, 
\A^{a},\Omega^{a}, \Sigma^{a}, \E^{a}, \H^{a}, \Sigma_{ab}, \E_{ab}, \H_{ab} }\,.
\ee 
Similarly, we may split the general energy momentum tensor in \rf{eq:einstScTneff} as:
\begin{equation}
T_{ab}=\mu u_a u_b +p h_{ab}+2q_{(a}u_{b)}+\pi_{ab}\;,
\label{eq:pf}
\end{equation}
where $\mu$ is the energy density  and  $p$ is the pressure.

The anisotropic fluid variables $q^{a}$ and $\pi_{ab}$ can be further split as:
\bea
q^{a} &=& Qe^{a} + Q^{a}~, \\
\pi_{ab} &=& \Pi\bras{ e_{a}e_{b} -\frac12 N_{ab}} + 
2\Pi_{(a}e_{b)} + \Pi_{ab}\;. \label{Anisotropic}
\eea

\subsection{The 1+1+2 equations for LRS-II Spacetimes} \label{Gen112Eq}
Because of its structure, the 1+1+2 formalism is ideally suited for a covariant description of all the Locally Rotationally Symmetric (LRS) spacetimes. These spacetimes possess a continuous isotropy group at each point \cite{EllisLRS} and exhibit locally a unique, preferred, covariantly defined spatial direction.  

Since LRS space-times are constructed to be isotropic, there are no preferred directions in the sheet and consequently all 1+1+2 vectors and tensors vanish. It follows that the only non-zero 1+1+2 variables are the  covariantly defined scalars  \cite{extension} $\brac{\A, \Theta,\phi, \xi, \Sigma,\Omega, \E, \H, \mu, p, \Pi, Q }.$

A subclass of the LRS spacetimes, called LRS-II, contain all the LRS spacetimes that are rotation free. As a consequence,  the variables $\Omega$, $ \xi $ and $ \H $ are identically zero in LRS-II spacetimes and 
$$\brac{\A, \Theta,\phi, \Sigma,\E, \mu, p, \Pi, Q }$$ fully characterise the kinematics. The propagation and constraint equations for these variables  are obtained by the Ricci and  (twice contracted) Bianchi identities and  can be found in \cite{extension}.

Note that the scalars given above characterise  any given space-time in a coordinate independent way and in this sense they are related to the Cartan scalars for this class of spacetimes \cite{Stephani}. This means that two exact solutions for which the 1+1+2 variables are different represent two different spacetimes. This is one on the main advantages of this method and it makes it worth studying, even in the simple case of static and spherically symmetric spacetimes.

Let us now turn to the case of spherically symmetric static spacetimes
which belong naturally to LRS class II. The condition of staticity implies that the dot derivatives of 
all the scalar quantities vanish. Therefore the expansion is zero, ($\Theta=0$), and this implies
$\Sigma=0$. This in turn yields, via the constraints, that the heat flux scalar $Q=0$. Hence the set of 1+1+2 equations which 
describe the spacetime become:
\bea \label{StSpSymEqGen}
\hat\phi =-\frac12\phi^2 -\frac23\mu-\frac12\Pi-\E\;,
\label{equation1a}\\
\hat\E -\frac13\hat\mu + \frac12\hat\Pi =- \frac32\phi\bra{\E+\frac12\Pi} \label{ConstrE}\;,
\label{equation2a}\\
0 = - \A\phi + \frac13 \bra{\mu+3p} -\E +\frac12\Pi\;,
\label{equation3a}\\ 
\hat p+\hat\Pi= -\bra{\frac32\phi+\A}\Pi-\bra{\mu+p}\A\;,
\label{equation4a}\\
\hat{\A\;} = -\bra{\A+\phi}\A + \frac12\bra{\mu +3p}\;,
\label{equation5a} \\
\hat{K} = -\phi K, \label{propGauss}\\
K = \frac13 \mu - \E - \frac12 \Pi + \frac14 \phi^{2}\;. \label{GaussCurv}
\eea
Eliminating $\E$ and using the constraints \rf{ConstrE} and  \rf{GaussCurv}   the system above can be reduced to
\bea \label{StSpSymEqGen}
&&\hat{\phi}=\mathcal{A} \phi -\mu -p-\Pi -\frac{\phi ^2}{2}\;,
\label{equation1b}\\
&&\hat{\A\;}=-\mathcal{A}\left(\mathcal{A}+ \phi \right)+\frac{1}{2} \left(\mu+3 p\right)\;,
\label{equation5b}\\
&&\hat{\Pi}+\hat{p}=-\mathcal{A} (\mu +p+\Pi )-\frac{3 \Pi  \phi
   }{2}\;,
\label{equation4b}\\
&& K = -p -  \Pi + \frac14 \phi (\A+\phi)\;. \label{GaussCurvb}\\
 && \E =\frac13 \bra{\mu+3p}+\frac12\Pi-\A\phi\;.\label{equation3b}
\eea
One could, of course, decide to eliminate other variables. In particular one might try to retain the \rf{propGauss}, due to its simplicity. However,  as we will see in Section \ref{SCTNEq}, this choice has to be taken with great care, especially when attempting to find exact solutions.

Once these equations have been solved it is useful to connect the 1+1+2 quantities to the metric coefficients and hence reconstruct the metric. Let us begin with the general spherically symmetric static metric:
\be\label{sphsymmmetric}
ds^{2}= - A(\rho)d\tau^{2} + B(\rho)d\rho^{2} + C(\rho)(d\theta^{2} + \sin^2\theta d\phi^{2})\;.
\ee   
Using the definition of covariant derivative one obtains:
\be
\A=e^a\dot{u}_a=\frac{1}{2A\sqrt{B}}\frac{d A}{d \rho} \;, 
\label{Asolution}
\ee
and
\be
\phi=\delta_a e^a=\frac{1}{C\sqrt{B}}\frac{d C}{d \rho}\;.
\label{Phisolution}
\ee
Note that we have two equations for three metric components, so at first sight it might seem that given a solution of the 1+1+2 potentials $\A$ and $\phi$, there is no way to determine the metric. One needs to remember, however, that the form of the coefficient $B$ depends on the choice of the coordinate $\rho$, so that the factor $\sqrt{B}$ can be reabsorbed into the definition of $\rho$ and effectively the metric \rf{sphsymmmetric} has only two unknown functions. Therefore a $p$ coordinate directly associated with the ``hat'' derivative,  would have $\hat{X}=X_{,p}$ which implies $B(p)=1$.

The results above reveal an interesting connection between the 1+1+2 formalism and the Takeno variables \cite{Takeno}. In fact one can see that many of the theorems proven by Takeno have a correspondence in the 1+1+2 framework.
\section{Spherically Symmetric Static Space-times in Scalar Tensor Gravity} \label{SCTNEq}
The simplest way to write the 1+1+2 equations for the case of ST gravity is to use the recasting of the field equations that given in Section II.  In particular $T^{\psi}_{ab}$ can be decomposed as in \rf{eq:pf} with
\begin{eqnarray}
\mu ^{\psi }&=&\frac{1}{F(\psi )}\left[ \frac{1}{2}\hat{\psi}^2+V+F' \left( \doublehat{\psi} + \phi \hat{\psi}\right)+ F'' \,\hat{\psi}^2\right]\;,\label{MuSig}\\
p^{\psi }&=&\frac{1}{F(\psi )}\left[-\frac{\hat{\psi}^2}{6}-V-\frac{2}{3} F'' \hat{\psi}^2-\frac{2}{3} F' \left(\doublehat{\psi}+\phi \hat{\psi}\right)-F'  \mathcal{A}\,\hat{\psi}\right]\;,\label{pSig}\\
\Pi^{\psi}&=&\frac{1}{F(\psi )}\left[\frac{2}{3}\hat{\psi}^2+\frac{2}{3} F'' \hat{\psi}^2+\frac{2}{3} F' \left(\doublehat{\psi}-\phi \hat{\psi}\right)\right]\;.\label{PiSig}
\end{eqnarray}
In this way it is possible to write (\ref{equation1a}-\ref{equation5a}) as 
\bea 
&& F \left[2 \hat{\phi}+(\phi -2 \mathcal{A})\phi\right]+2\hat{\psi}^2=2\mathcal{A} F' \hat{\psi} -2F''\hat{\psi}^2 -2 F' \doublehat{\psi}\;,\label{StSpSymSCTN1}\\
&&2 F \left[\hat{\A\;}+\mathcal{A} (\mathcal{A}+\phi )\right]+2 V= (3 \mathcal{A}+\phi ) F'\hat{\psi}-F'' \hat{\psi}^2-F' \doublehat{\psi}\;,
\label{StSpSymSCTN2}\\
&& \nn \hat{\psi}\left\{\left[3 \left(F'\right)^2+2 F\right] \doublehat{\psi}+F' \left(3 F''+1\right) \hat{\psi}^2\right.\nn\\&&~~~~~~~\left.+(\mathcal{A}+\phi ) \left[3 \left(F'\right)^2+2 F\right] \hat{\psi}+4 V F'-2 F V'\right\}=0\;, \label{eqDDs}
\label{StSpSymSCTN3}\\
&& K =\frac{\hat{\psi}^2}{2 F}+\frac{V}{F}+\frac{F'}{F} \hat{\psi}( \mathcal{A}+\phi) + \frac14 \phi (\A+\phi)\;, \label{StSpSymSCTN4}\\
&& \E =\frac{\hat{\psi}^2}{3 F}-\frac{2V}{3F}-\frac{F'}{F} \hat{\psi}( \mathcal{A}+\frac{1}{2}\phi) -\A\phi\;,\label{StSpSymSCTN5}
\eea
where we have assumed $F\neq0$. The above equations characterise the static and spherically symmetric solutions of a general ST theory of gravity. Note that in spite of the fact that the \rf{eqDDs}  contains second derivatives of the scalar field $\psi$,  it does not correspond exactly to the Klein-Gordon equation as we substituted the expression for $\hat{\phi}$ and $\hat{\A}$ in the third equation.  Note also that the case $3 \left(F'\right)^2+2 F=0$, which corresponds to the conformal coupling reduces the \rf{eqDDs} to 
\begin{equation}
\hat{\psi}\left\{2 \doublehat{\psi}+2(\mathcal{A}+\phi ) \hat{\psi}-2 V'\right\}=0\;,
\end{equation}
which is, as expected, the GR Klein-Gordon equation.

Let us now show how  (\ref{StSpSymSCTN1}-\ref{StSpSymSCTN5}) can be used to obtain exact solutions. The first thing that needs to be done is to choose a suitable radial coordinate.  A clever choice is to proceed in such a way that equation \rf{propGauss} has a trivial solution like in the case of  the coordinate $r$ for which 
$\hat{X}=-\frac{1}{2}r\phi\partial_rX$ in \cite{extension}. The Gauss curvature is therefore just $K=r^{-2}$. However, one must be careful in this respect to check that \rf{StSpSymSCTN4} is fulfilled, because the choice above decouples $K$ from $\phi$. This can be clearly seen by considering the theory
\bea
&&F=F_2 +\frac{2 F_1 e^{\frac{\phi_0 \psi }{2}}}{\phi_0}+F_2+\frac{2 \psi }{\phi_0},\\ &&V=-\frac{1}{4} \psi_0^2 \left[3 \phi_0 \left(4 F_1 e^{\frac{\phi_0 \psi }{2}}\phi_0+2 \psi \right)+10\right],
\eea
with the solution 
\be
\A=\frac{1}{2}\phi_0\;, ~~ \phi= \phi_0~~ \mbox{and}~~\psi=\psi_0 +2 \ln(\phi_0 r)
\ee
of (\ref{StSpSymSCTN1}-\ref{StSpSymSCTN3}), which corresponds to the following solution for the metric
\begin{eqnarray}
A=A_0 r^{2\frac{\phi_0}{|\phi_0|}}\;,~~ B= \psi_0^2r^2\;, ~~ C= r^{-2}\;.
\end{eqnarray}
Although the above solution satisfies (\ref{StSpSymSCTN1}-\ref{StSpSymSCTN3}) {\it and} \rf{propGauss}, it does not satisfy  \rf{StSpSymSCTN4} and is not a solution of the field equations \rf{eq:einstScTn}. This happens because the coordinate change removes the connection between $K$ and $\phi$ so that \rf{propGauss} does not guarantee that \rf{StSpSymSCTN4} is satisfied. 

On the other hand, the theory 
\be\label{ES1}
F=F_0 \psi^2 \qquad V= V_0 \psi^\beta
\ee
does satisfy the system (\ref{StSpSymSCTN1}-\ref{StSpSymSCTN4}) for this convenient choice of radial coordinate and it is easy to find the exact solution
\begin{eqnarray}
&& \A=\frac{\A_0}{r}\;,~~\phi=  \frac{2}{r}\\
&& K= \frac{ F_0 \left[2 \mathcal{A}_0 (\beta -4)+\beta-10\right]-2}{r^2(\beta -2)
   F_0}+\frac{ V_0 \sigma _0^{\beta }}{r^2F_0 \sigma_0^2},\\
&& \psi= \psi_0 r^{\frac{2}{\beta-2}}
\end{eqnarray}
which, in terms of the metric coefficients is given by:
\begin{eqnarray}\label{SolExact1}
A=r^{2\A_0}\;,~~ B=  1\;,~~ C= K^{-1}\;.
\end{eqnarray}
The above solution satisfies all the Einstein equations upon direct substitution. Since this solution does not reduce to Minkowski spacetime in any limit of the parameters, it is clearly not asymptotically flat.  The associated Newtonian potential can be calculated in the usual way and  contains a constant term of the same order of the gravitational constant\footnote{Upon conformal transformation $\tilde{g}_{ab}=\Omega^2 {g}_{ab}$
with $\Omega^2=F(\psi)$, the theory \rf{ES1} is mapped into general relativity with a minimally coupled scalar field with the potential
\begin{equation}
V(\psi)=e^{\frac{\sqrt{F_0} \psi }{\sqrt{6 F_0+1}}}.
\end{equation}
An exact solution for a similar theory has been found by Chan {\it t al.} in \cite{Chan:1995fr} and this means that the two solutions are related. Incidentally this solution is also related to the ones found in   \cite{Clifton:2005aj,Clifton:2006ug} that have been found in other contexts.}. 

\section{Some properties of the static and spherically symmetric metrics in Scalar Tensor gravity.}
Let us now use the 1+1+2 formalism we have developed in the first part of the paper to explore some of the properties of the static and spherically symmetric metrics in Scalar Tensor gravity. We will start with proving the absence of a Schwarzschild solution for these theories and the implications of this result for Birkhoff's theorem. We then discuss how one might characterise asymptotic flatness and quasi-local horizons.
\subsection{No Schwarzschild solution}
A first important question one can address using the system (\ref{StSpSymSCTN1}-\ref{StSpSymSCTN5}) is whether or not the theory  \rf{eq:actionScTn} admits in general a Schwarzschild solution. The reason is obvious: if we want to have a theory of gravity that is compatible with the constraints coming from celestial mechanics we will need to have that is like the one discovered by Schwarzschild or one that has similar characteristics\footnote{To be precise, one should add that the above statement is limited by the accuracy of our most precise measurements of the gravitational field in the Solar System and, of course, that are not considering any screening mechanism \cite{Babichev:2013usa}.}.

The Schwarzschild solution is obtained when $\phi$ and $\A$ satisfy
\begin{eqnarray}\label{schw}
\hat{\phi}+\frac{\phi ^2}{2}-\mathcal{A} \phi=0\;,~~\hat{\A\;}+\mathcal{A}^2+\mathcal{A}\phi=0\;.
\end{eqnarray}
In addition, since the Ricci scalar is identically zero, we find that the standard Klein-Gordon equation holds
\begin{equation} \label{}
\doublehat{\psi} +2 (\mathcal{A} - \phi)  \hat{\psi} -V'=0\;.
\end{equation}
Substituting the above equations in (\ref{StSpSymSCTN1}-\ref{StSpSymSCTN3}) and assuming $F\neq0$, we obtain:
\begin{eqnarray}
&&\hat{\psi}^2 \left(F''(\psi
   )+1\right)-\mathcal{A} \hat{\psi} F'(\psi )+ \doublehat{\psi} F'(\psi )=0\;,\\
&&\nn \doublehat{\psi}
   F'(\psi )+\hat{\psi} \left[\hat{\psi} F''(\psi )+(3
   \mathcal{A}+\phi ) F'(\psi )\right]\\&&~~~~~~~~~~~~~~~~~~~~~~~~~~+2 V(\psi
   )=0\;,\\
&&\hat{\psi} F'(\psi )
   \left[\hat{\psi} \left(2 (\mathcal{A}+\phi ) F'(\psi
   )-\hat{\psi}\right)+2 V(\psi )\right]=0,
\end{eqnarray}
It is easy to see that this system has a (double) solution, for $\hat{\psi}\neq 0$
\begin{equation} \label{}
F''(\psi)=-1\;,~~F'(\psi)=0\;,~~V(\psi)=\frac{1}{2}\hat{\psi}^2\;,
\end{equation}
which is clearly inconsistent. This means that the class of scalar tensor theories of gravity discussed in 
Section \ref{GenST} have {\it no} Schwarzschild solution if the scalar field is not trivial. This result implies that, 
in principle, the astrophysics and celestial mechanics in ST theories is different from what is obtained in GR. 

\subsection{The Birkhoff's  Theorem}
The conclusion above has an important consequences in terms of the meaning and role of the Birkhoff 
theorem in this class of theories. In this section we will use the 1+1+2 formalism to find out that although 
the originally Birkhoff theorem cannot be valid, there are other versions of this theorem that can be given for this classes of theories.

There are a number of formulations  and proof of this theorem in the context of General Relativity \cite{Faraoni,Hervik,Weinberg,Wald,Ohanian,Straumann,Stephani,Rindler,Shutz}. We will adopt the one presented in the book by Hawking and Ellis \cite{Hawking:1973uf} which can be stated as follows:
\begin{quote}
{\em Any $C^2$ solution of Einstein's equations which is static and spherically symmetric in an open set $\mathcal V$ is locally equivalent to part of the maximally extended Schwarzschild solution in  $\mathcal V$.}
\end{quote}
The  theorem can be easily proven in GR with the help of the 1+1+2 formalism. We will follow here the same approach of \cite{Ellis:2013dla} where a number of interesting cases is discussed in the framework of GR. The first step is to deduce the symmetry of the metric i.e.  its static and spherically symmetric character. We have already done this step naively in the previous section. We now derive these conditions rigorously.
 
The  Killing equation for a general vector $\xi_a$ orthogonal to the sheet is
\begin{equation}
\nabla_{(a}\xi_{b)}=0.
\end{equation}
We can always write $\xi_a$ as 
\begin{equation}
 \xi_a=\Psi \bar{u}_a+\Phi e_a+X_a, 
 \label{KV}
 \end{equation}
where we have chosen, for now, observers for which $\bar{u}_a \bar{u}^a=1$, $\Phi$ and $\Psi$ are scalars different form zero and $X_a=N_{ab}\xi^b$. Naturally from this definition one has  $\xi_a\xi^a=-\Psi^2+\Phi^2+X^aX_a
$, so that the character of $\xi_a$ depends on the values of $\Psi$ and $\Phi$ and the value of $X_a$ (i.e. $\Psi=0$ implies $\xi_a$ spacelike etc.). Decomposing the Killing equation and setting to zero the vector and tensor kinematical quantities, one obtains the following relations
\bea
\dot\Psi+\A\Phi&=&0, \label{psidot}\\
\hat\Psi -\dot\Phi-\Psi\A-\Phi\left(\frac\Theta3+\Sigma\right)+\dot{X}_b e^b+\hat{X}_b u^b
&=& 0,\label{psihat}\\
\hat\Phi+\Psi\left(\frac\Theta3+\Sigma\right)
&=&0,\label{cons1}\\
\Psi\left(\frac23\Theta-\Sigma\right)+\Phi\phi
&=&0,\label{cons2}\\
\delta_{\{a}X_{b\}}&=&0.
\eea 

Let us assume that $\xi^a$ is timelike and set $\Phi=0$ and coinsider only a Killing vectors in the $[u,e]$ 2-surface i.e. $X_a=0$. The non trivial Killing equations reduce to
\bea
\dot\Psi&=&0, \label{psidot1}\\
\hat\Psi +\Psi\A &=& 0,\label{psihat1}\\
\Psi\left(\frac23\Theta-\Sigma\right)&=&0,\label{cons11}
\eea 
The (\ref{psidot1}) and (\ref{psihat1}) always admit a solution so that the Killing vector exists and the \rf{cons11} gives $\Theta=0=\Sigma$. When these results are plugged into the general 1+1+2 equations, we reobtain the (\ref{StSpSymSCTN1}-\ref{StSpSymSCTN3}). This result implies  there always exists a Killing vector in the local $[u,e]$ plane for a vacuum LRS-II spacetime and therefore that the (\ref{StSpSymSCTN1}-\ref{StSpSymSCTN3}) represent all the solutions with a timelike Killing field.  In the case in which the Killing vector is spacelike one obtains  that $\A=0=\phi$ and we proceed in a similar manner, the difference being that in this case the solution will be homogenous. 
 
 In GR the  equations  corresponding to (\ref{StSpSymSCTN1}-\ref{StSpSymSCTN3}) lead to the Schwarzschild metric (exterior and interior), but here the solutions will be different. Suppose, for the moment, that such solution is known and let us refer to it as $\mathcal S$. The reasoning above tells us that if we have staticity and spherical symmetry in an open set $\mathcal V$ the metric of this solution is locally equivalent to part of the maximally extended of the $\mathcal S$ solution in  $\mathcal V$. 
 
Such result suggests a modification of  the Birkhoff theorem such as: 
\begin{quote}
{\em Any $C^2$ solution of (\ref{StSpSymSCTN1}-\ref{StSpSymSCTN3}) which is static and spherically symmetric in an open set $\mathcal V$ is locally equivalent to part of the maximally extended solution $\mathcal S$ in  $\mathcal V$.}
 \end{quote}
In order to prove the above statement we need only to prove that the solution $\mathcal S$ in unique. In order to do that  one has to prove that in the explicit form  the L.H.S. of (\ref{StSpSymSCTN1}-\ref{StSpSymSCTN3}) system is Lifschitz continuous  so that the  Picard-Lindel\"of theorem is satisfied. This implies that the scalar field $\sigma$, the functions $F$ and $V$ and their derivatives with respect to $\sigma$ need to be Lifschitz continuous. In addition, the condition 
\begin{equation} \label{LipschitzCondSys}
\hat{\sigma} F(\sigma) \left[2F(\sigma)+3F'(\sigma)^2\right]\neq 0,
\end{equation}
obtained solving (\ref{StSpSymSCTN1}-\ref{StSpSymSCTN3})  for $(\hat{\phi},\phi,\hat{\A},\A)$, needs to be satisfied\footnote{Note that because of the \rf{propGauss} in general  $K$ is Lifschitz continuous if $\phi$ has this property.}. This condition implies that $F$ should not have any zeros i.e. that the gravitational interaction cannot change sign. It is clear however than is $F$ admits zeros the solution can be still unique {\em within} these zeroes.

With these results one can therefore state the following modified Birkhoff theorem for Scalar Tensor gravity (mBT):
\begin{quote}
{\em Any $C^2$ solution of (\ref{StSpSymSCTN1}-\ref{StSpSymSCTN3}) which is static and spherically symmetric in an open set $\mathcal V$, for which $\sigma$ is  Lifschitz continuous and \rf{LipschitzCondSys} is satisfied, is locally equivalent to the maximally extended solution $\mathcal S$ in  $\mathcal V$.}
 \end{quote}
 which is holds  for any ST theory which satisfies \rf{LipschitzCondSys}.
 
The remaining problem therefore is to determine $\mathcal S$. The direct resolution of  (\ref{StSpSymSCTN1}-\ref{StSpSymSCTN3}) is however a formidable task. Thus it is worth looking for other ways to obtain the form of this solution.

One approach is to consider what has have learned in the case of $f(R)$-gravity. In  \cite{SSf(R)} it was found that the validity the original Birkhoff's theorem is guaranteed iff 
\begin{equation}
f(R)\big|_{R=0}=0, \qquad f'(R)\big|_{R=0}\neq0\;.
\end{equation}
Since we know that $f(R)$-gravity can be mapped into a Brans-Dicke-like theory with a non-trivial potential
\begin{eqnarray}
&&\psi=f'(R), \qquad R(\psi)=V'(\psi)  \label{SF-Transf1}
\\ &&V(\psi)=R(\psi)\psi-f(R(\psi)),  \label{SF-Transf2}
\end{eqnarray}
we can ask if the results of \cite{SSf(R)} provide any insight on the validity of the mBT for ST gravity. Unfortunately the answer is negative. In fact, since the Schwarzschild solution is characterised by $R=0$, (\ref{SF-Transf1}-\ref{SF-Transf2}) implies immediately that the scalar field is constant. In other words, the conditions found in \cite{SSf(R)} effectively correspond to GR via (\ref{SF-Transf1}-\ref{SF-Transf2}).

Another interesting way to attack this problem is to approach it from the point of view of conformal transformations. It is well known that under a conformal transformation ST theories of gravity of the type \rf{eq:actionScTn} are mapped into GR minimally coupled to a scalar field \cite{CF-Papers}. Can we then use the conformal transformations to find the form of the solution $\mathcal S$? 

In order to answer this question, let us consider the conformal transformation 
\begin{equation}\label{CT}
\tilde{g}_{ab}=\Omega^2 {g}_{ab}\;,
\end{equation}
with $\Omega^2=F(\sigma)$.
It is well known that under this transformation, equation \rf{eq:actionScTn} in vacuum can be recast as
\begin{equation}\label{eq:actionScTnConf}
\mathcal{A}=\int d x^{4}\sqrt{-g}\left[\frac{1}{2}R
-\frac{1}{2}\nab_a\varphi\nab^a\varphi -W(\varphi) \right]\;,
\end{equation}
where
\begin{equation} \label{DefPsi}
\varphi=\int \left( \frac{3 F'(\psi)^2-2F(\psi)}{2F(\psi)^2}\right)^{1/2} d\psi
\end{equation}
and 
\begin{equation} \label{PotPsi}
W(\varphi)=\left.\frac{V(\varphi)}{F(\varphi)}\right|_{\varphi=\varphi(\psi)}\;,
\end{equation}
As mentioned earlier, it has been recently shown that if the scalar field is static, the  (Jebsen-)Birkhoff theorem holds for these theories \cite{Faraoni}. In terms of the 1+1+2 approach a conformal transformation is composed by a boost and change of coordinates \cite{Carloni:2009gp}, so that 
\begin{eqnarray}
&\tilde{u}_a = \Omega u_a\\
&\tilde{e}_a = \Omega e_a\\
&\tilde{N}_{ab} = \Omega^2 N_{ab}
\end{eqnarray}
and the covariant derivative of a vector of conformal weight $s$ is transformed as
\begin{equation}
\tilde{\nabla}_{a}\tilde{V}_b=
{\nabla}_{a}(\Omega^s{V}_b) +\frac{\Omega^{s-1}}{2}\left(
2\delta^{a}_{(b} \nabla_{c)}\Omega -g_{bc}\nabla^{a} \Omega \right){V}_c
\end{equation}
It is straightforward to verify that a conformal transformation cannot change the symmetry of the spacetime. In fact under the conformal transformation above, the 1+1+2 vector quantities are mapped to zero. For example:
\begin{equation} \label{}
\tilde{\A_a}=\frac{{\A}_a}{\Omega}+\frac{\delta_a \Omega}{\Omega}\;.
\end{equation}
If the quantities on the RHS are subject to spherical symmetry, it is clear that $\tilde{\A_a}=0$. The same reasoning applies to all the other quantities. 

The Killing equations for $\xi_a$, which has $s=1$, in the Einstein frame become 
\begin{equation}
0= \tilde{\nabla}_{(a}\tilde{\xi}_{b)}=\Omega {\nabla}_{(a}{\xi}_{b)}+\frac{1}{2}g_{ab}\xi^c \Omega_{, c},
\end{equation}
which translates, assuming $\Phi=0$, $X_a=0$ and $\Omega\neq 0$, into
\bea
\dot\Psi-\frac{1}{2}\frac{\dot{\Omega}}{ \Omega }&=&0, \label{psidot1CFT}\\
\left(\hat\Psi -\Psi\A \right)\Omega &=& 0,\label{psihat1CFT}\\
\Psi\left(\frac23\Theta-\Sigma\right)+\frac{1}{2}\frac{\dot{\Omega}}{ \Omega }&=&0,\label{cons11CFT}
\eea 
It is clear therefore that a conformal transformation does not compromise the presence of a timelike Killing vector as far as the conformal factor is time independent (i.e. the metric remains static).  Therefore one can use conformal transformation to deduce the solution $\mathcal S$ analytically from a Birkhoff solution in the Einstein frame \footnote{In fact, the use of conformal transformations for the search of new solutions is known since long time \cite{Bekenstein}. The difference here is that we do not consider only a conformally invariant scalar field, relying on \rf{DefPsi} for the mapping of the solution of the scalar field. This step allows to extend the work done in \cite{Bekenstein} to general scalar tensor theories.}. For the theories for which this is possible the modified Birkhoff theorem above is valid.

Let us consider an example of such solution, which will be useful for what follows. Consider the minimally coupled theory\footnote{It is clear that, being this theory conformably related to the Brans-Dicke theory all the solutions given below are related conformally to the Brans solution. The real difference would arise if a non trivial potential was considered. As far as we are aware, no exact solution of this type (Birkhoff and asymptotically flat) has been presented in the literature. In the following we will use these solutions to illustrate how the 1+1+2 formalism is able to determine their properties.}
\begin{equation}\label{eq:actionSc}
\mathcal{A}=\int d x^{4}\sqrt{-g}\left[\frac{1}{2}R-\frac{1}{2}\nab_a\sigma\nab^a\sigma  \right]\;.
\end{equation}
The spherically symmetric solutions for these theory are well known \cite{Janis:1968zz,Virbhadra:1997ie,Wyman,Buchdahl}. A solution which is also asymptotically flat is given by
 \be\label{GRphiSol}
ds^{2}= - \tilde{A}(r)d\tau^{2} +  \tilde{B}(r)d\rho^{2} +  \tilde{C}(r)(d\theta^{2} + \sin^2\theta d\phi^{2})],
\ee   
where
 \bea
&& \tilde{A}(r)= \left(1-\frac{b}{r}\right)^\gamma,\\ 
&& \tilde{B}(r)= \left(1-\frac{b}{r}\right)^{-\gamma},\\
&& \tilde{C}(r)= \left(1-\frac{b}{r}\right)^{1-\gamma}r^2,
\eea   
with the  scalar field
\be
\psi=\sqrt{\frac{1-\gamma ^2} {2}}\log \left(1-\frac{b}{r}\right).
\ee
and $0<\gamma<1$. Using the results above we can generate a set of theories with accompanying solutions satisfying the mBT. 

For example, choosing 
\be
\Omega=\frac{1}{2 P^2}\left(1-\frac{b}{r}\right)^{\frac{P\sqrt{1-\gamma ^2} }{2 \sqrt{3}}}+\frac{1}{2}
   \left(1-\frac{b}{r}\right)^{-\frac{P\sqrt{1-\gamma ^2} }{2 \sqrt{3}}}\;,
\ee
or 
\be
\Omega=\frac{1}{2}\left(1-\frac{b}{r}\right)^{\frac{P\sqrt{1-\gamma ^2} }{2 \sqrt{3}}}+\frac{1}{2 P^2}
   \left(1-\frac{b}{r}\right)^{-\frac{P\sqrt{1-\gamma ^2} }{2 \sqrt{3}}}\;.
\ee
 We obtain:
\be
F(\psi)= \frac{1}{2 P^2}+\frac{1}{4} e^{-\sqrt{\frac{2}{3}}P}+\frac{e^{\sqrt{\frac{2}{3}} P \psi }}{4 P^4}+\frac{1}{4} e^{-\sqrt{\frac{2}{3}} P \psi},
\ee
and 
\be
F(\psi)=\frac{1}{2 P^2}+\frac{e^{-\sqrt{\frac{2}{3}} P \psi }}{4 P^4}+\frac{1}{4}e^{\sqrt{\frac{2}{3}} P \psi}\;,
\ee
respectively. In this way, the  coefficients of the accompanying solution satisfying the mBT are
\begin{eqnarray}
&&{A}= \frac{ \tilde{A}}{\Omega^{2}}=\frac{4 P^4 \left(1-\frac{b}{r}\right)^{\gamma +\frac{\sqrt{1-\gamma ^2}
   P}{\sqrt{3}}}}{\left[\left(1-\frac{b}{r}\right)^{\frac{\sqrt{1-\gamma ^2} P}{\sqrt{3}}}+P^2\right]^2}\label{SolCFT1}\\
&&{B}=\frac{  \tilde{B}}{\Omega^{2}}=\frac{4 P^4 \left(1-\frac{b}{r}\right)^{\frac{\sqrt{1-\gamma ^2} P}{\sqrt{3}}-\gamma
   }}{\left[\left(1-\frac{b}{r}\right)^{\frac{\sqrt{1-\gamma ^2} P}{\sqrt{3}}}+P^2\right]^2}\\
&&{ \tilde{C}}=\frac{ C}{\Omega^{2}}=\frac{4 P^4 r^2 \left(1-\frac{b}{r}\right)^{-\gamma +\frac{\sqrt{1-\gamma ^2}
   P}{\sqrt{3}}+1}}{\left[\left(1-\frac{b}{r}\right)^{\frac{\sqrt{1-\gamma ^2} P}{\sqrt{3}}}+P^2\right]^2\label{SolCFT3}}\\
\end{eqnarray}
with the scalar field solution
\be
\psi=\sqrt{\frac{1-\gamma ^2} {2P^2}}\log \left(1-\frac{b}{r}\right)\;.\label{SolCFT4}
\ee
In terms of the 1+1+2 parameters, this solution can be expressed as
\begin{eqnarray}
&\nonumber\A= \frac{b}{12 P^2 r^2}  \left(1-\frac{b}{r}\right)^{\frac{1}{6}\left(3( \gamma-2) -P\sqrt{3-3 \gamma ^2} \right)} \left[3 \gamma  \left(1-\frac{b}{r}\right)^{\frac{\sqrt{1-\gamma ^2} P}{\sqrt{3}}}\right.\\ &\left.-\sqrt{3-3 \gamma ^2} P
   \left(1-\frac{b}{r}\right)^{\frac{\sqrt{1-\gamma ^2} P}{\sqrt{3}}}+\sqrt{3-3 \gamma ^2} P^3+3 \gamma 
   P^2\right]\\
&\nonumber\phi=\frac{1}{6 P^2 r^2} \left(1-\frac{b}{r}\right)^{\frac{1}{6}\left(3( \gamma-2) -P\sqrt{3-3 \gamma ^2} \right)}   \left\{3 [b (\gamma +1)-2 r]
   \left(1-\frac{b}{r}\right)^{\frac{\sqrt{1-\gamma ^2} P}{\sqrt{3}}}\right.\\ 
 &\left.-b \sqrt{3(1- \gamma ^2)} P^3+3 P^2 [b (\gamma +1)-2 r]+b \sqrt{3-3 \gamma ^2} P
   \left(1-\frac{b}{r}\right)^{\frac{\sqrt{1-\gamma ^2} P}{\sqrt{3}}}\right\}
\end{eqnarray}
\subsection{ Asymptotic flatness} 
Another important property of physically relevant spherically symmetric solutions in general is asymptotic flatness. The general proof of this property for a given metric requires very refined theoretical tools such as  Penrose's method of conformal compactification \cite{bi:wald}. The covariant approach offers an interesting alternative, although a somewhat less general, approach to this problem. In fact decomposing the Riemann tensor in terms the 1+3 variables one obtains \cite{Covariant}
\begin{eqnarray}
R^{ab}{}_{cd} & = & R_P^{ab}{}_{cd} + R_I^{ab}{}_{cd} 
+ R_E^{ab}{}_{cd} + R_H^{ab}{}_{cd}\, , \\
R_P^{ab}{}_{cd} & = & \frac{2}{3}\,(\mu+3p)\,u^{[a}\,
u_{[c}\, h^{b]}{}_{d]} + \nn\\ && \frac{2}{3}\,\mu\,h^{a}{}_{[c}
\,h^{b}{}_{d]} \, ,\label{R_P}\label{R_P}\\ 
R_I^{ab}{}_{cd} & = & -\,2\,u^{[a}\,h^{b]}{}_{[c}\,q_{d]}
- 2\,u_{[c}\,h^{[a}{}_{d]}\,q^{b]} \nn
\\&&- 2\,u^{[a}\,u_{[c}\,\pi^{b]}{}_{d]} 
+ 2\,h^{[a}{}_{[c}\,\pi^{b]}{}_{d]} \,,\label{R_I}\\
R_E^{ab}{}_{cd} & = & 4\,u^{[a}\,u_{[c}\,E^{b]}{}_{d]} 
+ 4\,h^{[a}{}_{[c}\,E^{b]}{}_{d]} \,, \label{R_E} \\
R_H^{ab}{}_{cd} & = & 2\,\eta^{abe}\,u_{[c}\,H_{d]e} 
+ 2\,\eta_{cde}\,u^{[a}\,H^{b]e} \ . \label{R_H}
\end{eqnarray}
which in terms of the 1+1+2 variables and the static and spherically symmetric case reduces to 
\begin{eqnarray}
  R_P^{ab}{}_{cd} & = & \frac{2}{3}\,(\mu+3p)\,u^{[a}\,
u_{[c}\, h^{b]}{}_{d]} + \frac{2}{3}\,\mu\,h^{a}{}_{[c}
\,h^{b}{}_{d]} \, ,\\ 
  R_I^{ab}{}_{cd} & = & 
- 2\,u^{[a}\,u_{[c}\,\left(e^{b]}e_{d]}-\frac{1}{2}N^{b]}{}_{d]}\right)\Pi
+ 2\,h^{[a}{}_{[c}\,\left(e^{b]}e_{d]}-\frac{1}{2}N^{b]}{}_{d]}\right)\Pi \,,\\
  R_E^{ab}{}_{cd} & = & 4\,u^{[a}\,u_{[c}\,\left(e^{b]}e_{d]}-\frac{1}{2}N^{b]}{}_{d]}\right)\E
+ 4\,h^{[a}{}_{[c}\,\left(e^{b]}e_{d]}-\frac{1}{2}N^{b]}{}_{d]}\right)\E \,,\\
  R_H^{ab}{}_{cd} & = & 0\;.
\end{eqnarray}
If a metric is asymptotically flat, there will be a limit in which $u_a$, $e_a$ and $N_{ab}$ are constant tensors and the Riemann tensor is identically zero. This implies, by definition, that in this limit $\A$  and $\phi$ have to be zero and that the above relations become equations for $\mu$, $p$, $\Pi$ and $\E$. Using \rf{equation3a} it is easy to see that  
\begin{eqnarray}
\E & = & \frac13 \bra{\mu+3p}  +\frac12\Pi -\A\phi
\end{eqnarray}
which means
\begin{eqnarray}
\E \rightarrow\frac13 \bra{\mu+3p}  +\frac12\Pi\;.
\end{eqnarray}
It follows that the behaviour of $\E$ is determined by $\mu$, $p$ and $\Pi$: if these last quantities each tend asymptotically to zero then Riemann tensor will also tend to zero. Using (\ref{MuSig}-\ref{PiSig}), one obtains that, in the case of  \rf{eq:actionScTn}, this will occur if
\begin{equation}
\psi\rightarrow cost\;,~~ \frac{V(\psi)}{F(\psi)} \rightarrow 0,
\end{equation}
which is compatible with what is found in \cite{Sotiriou:2011dz}. It is interesting to note that in this limit \rf{equation1b} and \rf{equation5b} reduce to the equations that give rise to the Schwarzschild solution.

Let us now look at what happens to asymptotic flatness under a conformal transformation. It is easy to see that the relations (\ref{R_P}-\ref{R_H}) are invariant under conformal transformations. Therefore, the conditions for asymptotic flatness found from these equations remain the same. This is confirmed by the fact that the solution (\ref{SolCFT1}-\ref{SolCFT4}) remains asymptotically flat, as its conformal generator.

However, when matter is present, things are not as straightforward.  The thermodynamic quantities are rescaled via the conformal factor:
\begin{equation}
\tilde{\mu}= \Omega^2 \mu, \qquad \tilde{p}= \Omega^2 p, \qquad \tilde{\Pi}=\Omega^2 \Pi\;,
\end{equation}
therefore we require that if  a tilted thermodynamic quantity goes to zero this behaviour is guaranteed also for the un-tilded quantities. It is clear that this can happen only if the conformal factor asymptotically approaches a constant. 

In terms of the scalar field, these conditions amount to  
\begin{equation}
\Psi\rightarrow\Psi_0=const.\quad W(\Psi)=\frac{V(\Psi(\psi))}{F(\Psi(\psi))}\rightarrow 0. 
\end{equation}
 The first condition is satisfied only if \rf{DefPsi} converges to a constant when $F$ and its first derivative do so and this gives us a constraint on the ST theories that satisfy the mBT. 

\subsection{Horizons}
Let us conclude with the definition of quasi-local horizon in scalar tensor gravity in the language  of the 1+1+2 covariant approach. A black hole is usually characterised in terms of the presence of an event horizon and there are compelling reasons to do so \cite{DeWitt^2}. However in some cases it is useful to have a more local definition of black hole rather than one that depends on the global spacetime structure, which can be obtained using the definition of geometrically defined local horizons \cite{Nielsen}. This is particularly true in the case of scalar tensor theories in which the properties of black holes are not as well established as they are in GR. 

The nature of horizons, trapped surfaces and other similar features can be described using the flow of null geodesics (see e.g. \cite{Poisson,Wald,Faraoni:2013aba}). This definition falls naturally in the  framework of 1+1+2 approach. Here we limit ourselves to ones that are relevant to static spacetime i.e., trapped surfaces, anti-trapped surface, perfect horizon \cite{Hajicek} and Killing horizon.

Let us consider two congruences of null geodesics represented by the vector field $l_a$ and $\bar{l}_a$  ($l_a l^a=0=\bar{l}_a \bar{l}^a$) such that $\bar{l}_a \bar{l}^a=-\alpha$ where $\alpha>0$ is a constant. i.e. they present to congruences of null geodesics which flow in opposite directions. Following the standard procedure of the 1+1+2 approach we will decompose these vector fields as
\begin{equation}\label{nulled}
l_a=(-u_b l^b)(\kappa \tilde{u}_a+p_a), \qquad \bar{l}_a=(-u_b \bar{l}^b)(\bar{\kappa} \tilde{u}_a+\bar{p}_a).
\end{equation}
where $\tilde{u}_a$ is obtaining normalising the time-like vector $u_a$ so that, in $l_a$ for example, $u_a=\kappa \tilde{u}_a$.  If these null geodesics are associated with a ingoing or an outgoing flux of electromagnetic waves in geometric optic approximation, the quantity $-u_b {l}^b=E$ and $-u_b \bar{l}^b=\bar{E}$ will be proportional to the  photon frequencies as measured by the observers $u_a$ and
\begin{equation}
p_a=\frac{h_{a}^bl_b}{E}, \qquad \bar{p}_a=\frac{h_{a}^b\bar{l}_b}{\bar{E}}.
\end{equation}
 are the propagation directions of the ``rays''.
  
Next, we can construct the 2-surface :
\begin{equation}
s_{ab}=\alpha g_{ab}+2 {l}_{(a}\bar{l}_{b)}.
\end{equation}
so  that  $s_{ab}l^b=0$  and  $s_{ab}\bar{l}^b=0$ i.e. $s_{ab}$ orthogonal to both $l_a$ and $\bar{l}_a$. At this point the expansion of the null geodesics through the surfaces associated to $s_{ab}$ are
\begin{eqnarray}
&& \Theta^+=s_{ab}\nabla^a l^b=\nabla^a l_a\\
&&  \Theta^-=s_{ab}\nabla^a \bar{l}^b=\nabla^a \bar{l}_a.
\end{eqnarray}
The vectors $p_a$ and $\bar{p}_a$ can be further split extracting the component along $e_a$:
\begin{equation}
p_a=\varepsilon e_a+ q_a, \qquad \bar{p}_a=\bar{\varepsilon}  e_a+ \bar{q}_a.
\end{equation}
Here, differently from the previous sections, we put ourselves in the general situation in which $e_a$ is not normalized and we suppose $e_ae^a=\eta$. In the case of LRS-II spacetimes one obtains, for the expansion of the rays $l_a$,
\begin{eqnarray}
\Theta^+=\varepsilon \left( \kappa^2\A E+\hat{E} \eta +\eta \hat{\eta} + E\phi\right),
\end{eqnarray}
By definition the vector field will satisfy the relations $l^a\nabla_a l_b=0$ and $l^b\nabla_a l_b=0$. The component of  the first of these equations along $u_a$  implies, in the case of LRS-II spacetimes
\begin{eqnarray}
\hat{E}\eta \kappa + E \eta\A \kappa+ E\hat{\kappa}=0.
\end{eqnarray}
Substituting in $\Theta^+$, one obtains
\begin{eqnarray}
\Theta^+= \varepsilon E\left[\eta^2 \left(-\frac{\hat{\kappa}}{\kappa}+\frac{\hat{\eta}}{\eta}\right) +\phi\right].
\end{eqnarray}
The same reasoning can be applied to the derivation of $\Theta^-$ to obtain
\begin{eqnarray}
\Theta^-= \bar{\varepsilon} E\left[\bar{\eta}^2 \left(-\frac{\hat{\bar{\kappa}}}{\bar{\kappa}}+\frac{\hat{\bar{\eta}}}{\bar{\eta}}\right) +\phi\right].
\end{eqnarray}
Since the values of $\kappa$, $\bar{\kappa}$, $\eta$ and $\bar{\eta}$ depend on the choice of the observer, the  above equation tells us that if in an event $\phi=0$ there exist observers for which $\Theta=0$ at that event\footnote{From this expression, in principle for observers which have $\eta^2 \left(-\frac{\hat{\kappa}}{\kappa}+\frac{\hat{\eta}}{\eta}\right) +\phi=0$ one can have $\phi\neq0$ and $\Theta=0$ at all time. This special class of observers would never see light rays converge or expand. In the following, we will never use such observers, relying only on the natural parameter $p$ and the area radius $r$, so the condition $\phi =0$ can be used without problems.}. In other words $\phi =0$ is a necessary condition for the existence of observers for which $\Theta=0$ is zero. For some specific choices of the affine parameter $\phi =0$ is also a sufficient condition. In the case of the area radius $r$, for example, one has
\begin{eqnarray}
\Theta^+= \frac{1}{2} \varepsilon E\phi \left[r \eta^2\left(\frac{\eta_{,r}}{\eta}-\frac{\kappa_{,r}}{\kappa}\right) +2\right],\\
\Theta^-= \frac{1}{2} \bar{\varepsilon} \bar{E}\phi \left[r \bar{\eta}^2\left(\frac{\bar{\eta}_{,r}}{\bar{\eta}}-\frac{\bar{\kappa}_{,r}}{\bar{\kappa}}\right) +2\right].
\end{eqnarray}
According to the definitions in \cite{Gourgoulhon:2008pu,Booth:2005qc,Nielsen:2008cr}, we can say, therefore, that the sign of $\phi$ gives a necessary condition for an observer to see a trapped surface. More specifically
\begin{equation}
\left\{ \begin{array}{ll}
&\phi<0 \rightarrow \mbox{trapped surfaces}\\
&\phi>0 \rightarrow \mbox{anti-trapped surfaces}
\end{array}\right.
\end{equation}
and $\phi=0$ will give the ``non expanding horizon'' or``perfect horizon'' \cite{Hajicek}. 

Another important type of horizon is the so called Killing horizon. A Killing horizon is defined as the surface in which the modulus of a Killing vector of the metric is zero. A spacetime which is static and spherically symmetric contains four killing vector: a timelike one, which is associated to the static nature of the metric (in fact the Killing vector only guarantees only guarantees whether the metric is stationary) and other three which represent spherical symmetry. 

It is relatively easy to characterise the vanishing of the modulus of the timelike Killing vector. Setting again $\Phi=0$  and $X_a=0$ in \rf{KV}, we obtain from \rf{psihat1} that, formally,
\begin{equation}
\log\Psi=-\int_{\mathcal Q}\A d p
\end{equation}
where ${\mathcal Q}$ is the domain in which the metric is static and spherically symmetric\footnote{To be consistent with the treatment of perfect horizon, we should derive this relation in the case in which $u_a$ is not normalised. However this would lead to a relation of the type 
\begin{equation}
\log\left(\Psi \kappa\right)=-\int_{\mathcal Q}\A d p
\end{equation}
where $\kappa$ is the modulus of $u_a$. We will assume in the following that $\Psi=\kappa \Psi$ i.e. $\Psi$ contains already the contribution of the normalization.}. This implies that a Killing horizon is only realised when 
\begin{equation}\label{KillingIntegral}
\int_{\mathcal Q}\A d p\rightarrow \infty.
\end{equation}
If ${\mathcal Q}$ is a finite interval the above result means  $\A$ is divergent at some point  in ${\mathcal Q}$. If ${\mathcal Q}$ is of the form $] a, \infty[$ then the above relation might imply that the metric is not  asymptotically flat. Of course the Killing horizon will be present only in the first case. It is important to keep in mind that, as in the case of the perfect horizon, the presence of an horizon is a coordinate dependent concept and therefore the outcome of the integral above depends also on the choice of the affine parameter. In the case of the area radius one has
 \begin{equation}
\int_{\mathcal Q(p)}\A d p=\int_{\mathcal Q(r)} \frac{2 \A}{r \phi}d r,
\end{equation}
therefore in this coordinate system one requires the fact that $\A/\phi$ is divergent in the domain of integration and therefore a $\phi$ that admits a zero can generate a Killing horizon also when $\A$ is regular. In other words, one can say that a divergence in $\A$ is a sufficient condition for the presence of a Killing horizon\footnote{What about the Killing vectors on the sheet? Since the presence of a Killing horizon is a coordinate dependent statement, in this case the 1+1+2 covariant approach (because it is a partial tetrad) only tells us that these Killing vectors will be the solutions of the one associated with the topology of the sheet (closed, flat, open) and we will need to choose a set of coordinates (or a full tetrad) to determine their presence. However in the static spherically symmetric case we know that there is no horizon corresponding to these Killing vectors in the standard coordinate systems used to investigate the properties of these solutions.}.

Let us consider as an example the standard Schwarzschild solution in GR. In terms of the area radius one has
\begin{equation}
\A=\frac{m}{r^2}\left(1-\frac{2m}{r}\right)^{-1/2}\quad \phi =\frac{2}{r}\left(1-\frac{2m}{r}\right)
\end{equation}
At the horizon $\A$ is already divergent, however for this choice of the affine parameters the  Killing horizon is determined by the divergence of the quantity $\A/\phi$ which happens at the same value of $r$. Since in this case $\phi=0$ is a necessary and sufficient condition for the existence of a perfect horizon, it is clear that in this case  the perfect horizon is also a Killing horizon. 

The relation between horizons and singularities can be explored using the \rf{StSpSymSCTN4} or the constraint on the scalar component of the electric part of the Weyl tensor \rf{equation3b}. In the case of  $F=const.$ and in vacuum:
\begin{equation}
K =\frac14 \phi (\A+\phi)\;.
\end{equation}
and
\begin{equation}
\E = -\A\phi\;.
\end{equation}
Therefore when the Killing horizon is generated by a divergence in $\A$, the presence of a perfect horizon prevents the appearance of a singularity at the horizon. This is exactly what happens in the Schwarzschild case on the Schwarzschild radius. If only a perfect horizon is present, then  $\A$ is regular and $K$ and $\E$ are identically zero at the horizon. If, instead, only a Killing horizon is present, like in $r=0$ in the Schwarzschild case, it will lead to a singularity.

The presence of a matter term (regular or effective) changes the situation. More specifically, the term  $-(p + \Pi)$  appears in the constraint \rf{GaussCurvb} and the term $\frac{1}{3}(\mu+ 3p)$ in the \rf{equation3b}. This implies that (effective) matter can introduce further singularities or ``renormalise'' the ones associated to a Killing horizon. For example, a divergence of $\A$ can be compensated by a divergence in these term. In this case, therefore, perfect horizon and Killing horizon can be different without implying the presence of an actual singularity of the spacetime. 

Consider, for example, the solution \rf{GRphiSol}. The 1+1+2 potentials are given by
\begin{eqnarray}
&&\A=\frac{b \gamma  \left(1-\frac{b}{r}\right)^{\frac{\gamma }{2}-1}}{2 r^2}\\
&&\phi=\frac{2}{r} \left(1-\frac{b}{r}\right)^{\frac{\gamma }{2}-1} \left(1-\frac{b (\gamma +1)}{2
   r}\right)
\end{eqnarray}
The condition $\phi=0$ gives two perfect horizons: $r=b$ and $r=\frac{b}{2}(1+\gamma)$. The quantity  $\A$ is divergent for $r=b$, but in the coordinate system of the metric the integrand of \rf{KillingIntegral} is given by
\begin{equation}
 \A \sqrt{B}=-\frac{b \gamma }{2 r (b-r)}
\end{equation}
which has only one finite pole when $r=b$. Therefore the metric admits a Killing and perfect horizon for $r=b$ and another perfect horizon in $r=\frac{b}{2}(1+\gamma)$. Note that $\E$ and $K$ both diverge in $r=b$, so this horizon constitute a singularity for the spacetime. This is due to the presence of the kinetic term for the scalar field in the expression of $\E$ and $K$, as explained above. These two last quantities  are instead regular in $r=\frac{b}{2}(1+\gamma)$. This implies that the perfect horizon in $r=\frac{b}{2}(1+\gamma)$ can be eliminated by a coordinate transformation.

The case of ST gravity is even more complex that the non vacuum GR case.  In the  \rf{StSpSymSCTN4}and \rf{StSpSymSCTN5} this is evident from the appearance of the term $\frac{\hat{F}}{F} $ coupled with two different combinations of $\mathcal{A}$ and $\phi$. These terms show the effect of the non minimal coupling in the relation between Killing horizon, perfect horizon and singularities.   If a ST theory presents a Killing horizon  associated to a divergence of $\A$ the  \rf{StSpSymSCTN4} and \rf{StSpSymSCTN5} shows that the existence of a perfect horizon will not be sufficient anymore to avoid the presence of a singularity. Conversely a divergence of, say, the scalar field at the horizon will not translate in an actual divergence of the spacetime. 

It is instructive to compare our conclusion with the structure of some exact solutions. In the case of \rf{SolExact1} we have 
\begin{equation}
\A=\frac{\A_0}{r}\;,~~\phi=  \frac{2}{r}
\end{equation}
so no Killing or perfect horizons are present. 

Let us now consider the  solution found by Bekenstein \cite{Bekenstein} and Bocharova, Bronnikov and Melnikov (BBM) \cite{BBM} (in the case of zero electric field):
\be
ds^{2}= - {A}(r)d\tau^{2} +  {B}(r)d\rho^{2} +  r^2(d\theta^{2} + \sin^2\theta d\phi^{2})],
\ee   
where
 \bea
&& A(r)= \left(1-\frac{R_0}{r}\right)^2,\\ 
&& {B}(r)= \left(1-\frac{R_0}{r}\right)^{-2},\\
\eea  
with the  scalar field
\be
\psi=\frac{\psi_0}{\left(r-R_0\right)}.
\ee 
For this metric in terms of the coordinate $r$ we have 
\begin{equation}
\A=\frac{R_0}{r^2}\;,~~\phi=  \frac{2}{r^2}(r-R_0)
\end{equation}
From our previous reasoning  $r=R_0$ is a perfect horizon but $\A$ is regular and therefore the Killing horizon is not connected to any singularity of the metric. It is known, however that the scalar field is  singular on the horizon and the  \rf{StSpSymSCTN4} and \rf{StSpSymSCTN5} show clearly that this is due to compensation between the divergence of  function $F$ and the first derivative of the scalar field.

Finally let us look at (\ref{SolCFT1}-\ref{SolCFT4}). Because of the domain of the transformation the metric is only defined for $r>b$  and in this range $\phi\neq0$ and 
 \begin{equation}
 \int_{r>b} \frac{2 \A}{r \phi} d r<\infty,
\end{equation}
 there is no perfect or Killing horizon. Since the transformation that generates (\ref{SolCFT1}-\ref{SolCFT4})  is only valid for $0<\gamma<1$ the perfect horizon appearing in the Einstein frame for $r=\frac{b}{2}(1+\gamma)$ does not appear in this solution\footnote{Note, however, that  since this point is on the very edge of the domain of the conformal transformation such conclusion should be supported by further exploration of the properties of the conformal transformation. We will not address this issue here.}. Note that the nature of the spatial surface $r=b$ can be also modified by the transformation: for $P<\frac{\sqrt{3}(\gamma-2)}{\sqrt{\gamma-1}}$ it is not  singular anymore.

\section{Conclusions.}  
In this paper we have used the 1+1+2 formalism to analyse the spherically symmetric metrics in non-minimally coupled ST gravity. As in the case of the 1+3 covariant approach, our method can be easily applied if one treats the non-Einsteinian parts of the gravitational interaction as an effective fluid. 

The key 1+1+2 equations form a closed system of three differential equation together with two constraints, which can be used to find new exact solutions. However, great care should be taken in choosing the radial coordinate, as  some of these choices can result in a decoupling of the key equations, leading to solutions of the 1+1+2 equations which are not solutions of the full Einstein equations.

An interesting result of this paper relates to the existence of the Schwarzschild solution in ST gravity and on how this impacts on the original formulation of Birkhoff's theorem. Using the 1+1+2 equations it is easy to show in a coordinate independent manner that no ST theory admits a Schwarzschild solution unless the scalar field is trivial. It follows that one cannot define a Birkhoff theorem in the usual way. One can, however,  propose an extension to this theorem (mBT) in which the role of the Schwarzschild solution is taken by the general static and spherically symmetric solution for these theories. The problem then arises to determine the existence of this solution. 

A possible way to address this issue can be found using the well known results on the conformal symmetry of vacuum solution for ST gravity. Since the 1+1+2 formalism offers a relatively straightforward way to understand  the conditions under which the Birkhoff theorem is preserved under conformal transformations, one can use these prescriptions to generate theories and solution that comply the mBT. In this way the problem of the existence of  mBT complying solution is reduced to the problem of finding solution in the minimally coupled case which satisfies the Birkhoff theorem. 

The 1+1+2 formalism also allows a clear description of two important physical properties of these metric, namely the asymptotic flatness and the presence of quasi local horizons. In both cases the 1+1+2 returns some coordinate independent and physically clear conditions, which can be used to characterise the properties static spherically symmetric solutions not only of scalar tensor theories but also of any other extension of general relativity. Our results show that the use of this formalism can, as  has already been shown in cosmology, help clarify some of the unresolved issues in the physics of LRS-II spacetimes beyond General Relativity.

\section*{ Acknowledgments} 
We wish to thank V. Faraoni and T. Clifton for useful comments. SC  was supported by  the Funda\c{c}\~{a}o para a Ci\^{e}ncia e Tecnologia through project IF/00250/2013 and partly funded through H2020 ERC Consolidator Grant - Matter and strong-field gravity: New frontiers in Einstein's theory- grant agreement no. MaGRaTh-64659. PKSD thanks the NRF (South Africa) for financial support. 


\begin{thebibliography}{999}
\bibitem{bi:BransDicke}
C. Brans and H. Dicke,
Phys. Rev.  {\bf 124}, 925 (1961)
\bibitem{LaStein} D. La,  P. J.  Steinhardt, 
Phys. Rev. Let. {\bf 62(4)}, 376-378 (1989).
\bibitem{ScTnDarkEnergy} See for example
 E.~Elizalde, S.~Nojiri and S.~D.~Odintsov,
Phys.\ Rev.\ D {\bf 70} (2004) 043539
[arXiv:hep-th/0405034].
\bibitem{Donoghue} J. F. Donoghue, Phys. Rev. {\bf D50}, 3874 (1994).
\bibitem{Bekenstein}  	
J. D. Bekenstein, 
in Cosmology and Gravitation, M. Novello, ed. (Atlantisciences, France 2000), pp. 1-85,(1998)  arXiv preprint gr-qc/9808028.
\bibitem{Bergmann}P. G. Bergmann, Annals of Mathematics, {\bf49}, 255Ð264 (1948)
\bibitem{Fuji-Maeda} Y. Fujii, K. Maeda {\em The scalar-tensor theory of gravitation} Cambridge University Press, (2003).
\bibitem{FaraoniBook} V. Faraoni, {\em Cosmology in scalar-tensor gravity} Kluwer Academic Publishing (2004).
\bibitem{Damour:1993hw}
  T.~Damour and G.~Esposito-Farese,
  Phys.\ Rev.\ Lett.\  {\bf 70} (1993) 2220.
\bibitem{Bekenstein} J. D. Bekenstein,  
  Ann. of Phys. {\bf82.2} 535 (1974)
\bibitem{GRLimit} T. Damour, K. Nordtvedt  Phys. Rev. Lett. {\bf70}, 2217Ð2219 (1993)   ; J. P. Mimoso, A. Nune,  Astrophysics and Space Science, {\bf26}, 327
\bibitem{Brans} C.H. Brans, Phys. Rev. {\bf 125}, 2194 (1962); A. Bhadra, K. Sarkar, Gen. Relat. Gravit. {\bf 37}, 2189 (2005). See also E.~Babichev, C.~Charmousis and A.~Lehébel,
  Class.\ Quant.\ Grav.\  {\bf 33} (2016) no.15,  154002
  doi:10.1088/0264-9381/33/15/154002
  [arXiv:1604.06402 [gr-qc]], A.~Maselli, H.~O.~Silva, M.~Minamitsuji and E.~Berti,
  Phys.\ Rev.\ D {\bf 93} (2016) no.12,  124056
  doi:10.1103/PhysRevD.93.124056
  [arXiv:1603.04876 [gr-qc]],  and  E.~Babichev, K.~Koyama, D.~Langlois, R.~Saito and J.~Sakstein,
  arXiv:1606.06627 [gr-qc] and references therein for the case of Horndeski theories.
\bibitem{Covariant}
J.  Ehlers Abh. Mainz Akad. Wiss. u. Litt. (Math. Nat. kl) 11 (1961);
G. F. R. Ellis, in General Relativity and Cosmology, Proceedings of
XLVII Enrico Fermi Summer School, ed . R. K, Sachs (New York Academic
Press, 1971);
G. F. R. Ellis and H. van Elst 
Cosmological models (Carg\`{e}se lectures 1998),
in {\em Theoretical and Observational Cosmology\/}, edited by
M. Lachi\`{e}ze-Rey, p. 1 (Kluwer, Dordrecht, 1999).
\bibitem{ellisbook} Wainwright J and Ellis G F R (ed) 1997 {\it
Dynamical systems in cosmology} Cambridge: Cambridge University
Press (see also references therein)
\bibitem{SanteST}S.~Carloni, S.~Capozziello, J.~A.~Leach and P.~K.~S.~Dunsby,
  Class.\ Quant.\ Grav.\  {\bf 25} (2008) 035008
  [gr-qc/0701009]; S.~Carloni, P.~K.~S.~Dunsby and C.~Rubano,
  Phys.\ Rev.\ D {\bf 74} (2006) 123513
  [gr-qc/0611113].
\bibitem{extension}
C.~A.~Clarkson and R.~K.~Barrett,
Class.\ Quant.\ Grav.\  {\bf 20}, 3855 (2003)
[arXiv:gr-qc/0209051];
C.~Clarkson,
Phys.\ Rev.\ D {\bf 76}, 104034 (2007)
[arXiv:0708.1398 [gr-qc].
 C.~A.~Clarkson, M.~Marklund, G.~Betschart and P.~K.~S.~Dunsby,
 Astrophys.\ J.\  {\bf 613}, 492 (2004)
 [astro-ph/0310323].
 \bibitem{LensingST} C.~Schimd, J.~P.~Uzan and A.~Riazuelo,
  Phys.\ Rev.\ D {\bf 71} (2005) 083512
  [astro-ph/0412120]; V.~Faraoni and E.~Gunzig,
  Astron.\ Astrophys.\  {\bf 332} (1998) 1154
  [astro-ph/9801172; K.~Sarkar and A.~Bhadra,
  Class.\ Quant.\ Grav.\  {\bf 23} (2006) 6101
  [gr-qc/0602087];T.~Narikawa, T.~Kobayashi, D.~Yamauchi and R.~Saito,
  Phys.\ Rev.\ D {\bf 87} (2013) 124006
  [arXiv:1302.2311 [astro-ph.CO]].
\bibitem{Betschart}
C. A. Clarkson, M. Marklund,  G. Betschart and P. K. S. Dunsby, 
2003 Astrophys.J. 613 492-505 (2004).
  \bibitem{Nordtvedt} K. Nordtvedt 
Astrophys. J., {\bf161}, 1059 (1970); P. G. Bergmann,  Int. J. of Th. Phys. {\bf(1)} 25-36  (1968)
  \bibitem{bi:wald}
Wald, R. M. 1984, {\it General Relativity} (The University of
Chicago Press).
\bibitem{Galileon} Nicolis A, Rattazzi R and Trincherini E  Phys. Rev. D {\bf 79} 064036 (2009); Fairlie D B and Govaerts J  J. Math. Phys. 33 3543?66 (1992)
\bibitem{Horndeski}  Horndeski G W,  Int. J. Theor. Phys. {\bf10} 363?84 (1974)
\bibitem{EllisLRS} H. van Elst, G.F.R. Ellis 
 CGQ{\bf13.5}, 1099 (1996)
 \bibitem{Takeno} H. Takeno, "The Theory of Spherically Symmetric Space-Times" (Daigaku, Hiroshima, Japan, 1963).
 \bibitem{Chan:1995fr}
  K.~C.~K.~Chan, J.~H.~Horne and R.~B.~Mann,
  Nucl.\ Phys.\ B {\bf 447} (1995) 441
  [gr-qc/9502042].
\bibitem{Clifton:2006ug}
  T.~Clifton,
  Class.\ Quant.\ Grav.\  {\bf 23} (2006) 7445
  [gr-qc/0607096].
\bibitem{Clifton:2005aj}
  T.~Clifton and J.~D.~Barrow,
  Phys.\ Rev.\ D {\bf 72} (2005) 103005
  [gr-qc/0509059].
\bibitem{Babichev:2013usa}
  E.~Babichev and C.~Deffayet,
  Class.\ Quant.\ Grav.\  {\bf 30} (2013) 184001
  doi:10.1088/0264-9381/30/18/184001
  [arXiv:1304.7240 [gr-qc]].
 \bibitem{Faraoni} V.~Faraoni,
  Phys.\ Rev.\ D {\bf 81} (2010) 044002
  [arXiv:1001.2287 [gr-qc]].
  \bibitem{Carloni:2009gp} 
  S.~Carloni, E.~Elizalde and S.~Odintsov,
  Gen.\ Rel.\ Grav.\  {\bf 42}, 1667 (2010)
  [arXiv:0907.3941 [gr-qc]].
 \bibitem{Hervik} G. \O yvind, S. Hervik. {\it Einstein's general theory of relativity: with modern applications in cosmology} Springer Verlag, 2007.
 \bibitem{Weinberg} S. Weinberg {\it Gravitation and cosmology: principles and applications of the general theory of relativity} Vol. 1. New York: Wiley, 1972.
 \bibitem{Wald} R. M. Wald, {\it General relativity} University of Chicago press, 2010.
 \bibitem{Ohanian} H. C. Ohanian,  R. Ruffini  {\it Gravitation and spacetime} New York: Norton, 1976.
 \bibitem{Straumann} N. Straumann,  {\it General relativity and relativistic astrophysics}, Berlin and New York, Springer-Verlag (1984).
 \bibitem{Stephani} H. Stephani, ed. Exact solutions of Einstein's field equations. Cambridge University Press, 2003.
 \bibitem{Rindler} W. Rindler, {\it Relativity: special, general, and cosmological}  Oxford University Press, Oxford, UK
 \bibitem{Shutz} B. Schutz, {\it A first course in general relativity} Cambridge university press, 1985.
 \bibitem{Hawking:1973uf}
  S.~W.~Hawking and G.~F.~R.~Ellis,
 {\it The Large Scale Structure of Space-Time},  Cambridge university press, 1973.
\bibitem{Ellis:2013dla}
  G.~F.~R.~Ellis and R.~Goswami,
  Gen.\ Rel.\ Grav.\  {\bf 45} (2013) 2123
  [arXiv:1304.3253 [gr-qc]];
  R.~Goswami and G.~F.~R.~Ellis,
  Gen.\ Rel.\ Grav.\  {\bf 44} (2012) 2037
  [arXiv:1202.0240 [gr-qc]];
  R.~Goswami and G.~F.~R.~Ellis,
  Gen.\ Rel.\ Grav.\  {\bf 43} (2011) 2157
  [arXiv:1101.4520 [gr-qc]].
 \bibitem{Sotiriou:2007zu}
  T.~P.~Sotiriou, V.~Faraoni and S.~Liberati,
  Int.\ J.\ Mod.\ Phys.\ D {\bf 17} (2008) 399
  [arXiv:0707.2748 [gr-qc]].
  \bibitem{Sotiriou:2011dz}
  T.~P.~Sotiriou and V.~Faraoni,
  Phys.\ Rev.\ Lett.\  {\bf 108} (2012) 081103
  [arXiv:1109.6324 [gr-qc]].
 \bibitem{SSf(R)}
A.~M.~Nzioki, S.~Carloni, R.~Goswami and P.~K.~S.~Dunsby,
  Phys.\ Rev.\ D {\bf 81} (2010) 084028
  [arXiv:0908.3333 [gr-qc]].
\bibitem{CF-Papers}
 V.~Faraoni, E.~Gunzig and P.~Nardone,
  Fund.\ Cosmic Phys.\  {\bf 20} (1999) 121
  [gr-qc/9811047]; S.~Carloni, E.~Elizalde and S.~Odintsov,
  Gen.\ Rel.\ Grav.\  {\bf 42} (2010) 1667
  [arXiv:0907.3941 [gr-qc]].
 \bibitem{DeWitt^2} see e.g. C. DeWitt, B.S. DeWitt (eds.) {\it BlackHoles,Les astres occlus}, Gordon and Breach Science Publishers, NY, USA (1973) .
  \bibitem{Nielsen} A.B. Nielsen
Gen. Relativ. Gravit. {\bf 41}, 1539-1584 (2009)
\bibitem{Poisson}  E. Poisson {\it A Relativist's Toolkit: The Mathematics of Black-Hole Mechanics}, Cambridge University Press: Cambridge, UK, 2004.
\bibitem{Faraoni:2013aba}
  V.~Faraoni,
  Galaxies {\bf 1} (2013) 3,  114
  [arXiv:1309.4915 [gr-qc]].
\bibitem{Gourgoulhon:2008pu}
  E.~Gourgoulhon and J.~L.~Jaramillo,
  New Astron.\ Rev.\  {\bf 51} (2008) 791
  [arXiv:0803.2944 [astro-ph]].
\bibitem{Booth:2005qc}
  I.~Booth,
  Can.\ J.\ Phys.\  {\bf 83} (2005) 1073
  [gr-qc/0508107].
\bibitem{Nielsen:2008cr}
  A.~B.~Nielsen,
  Gen.\ Rel.\ Grav.\  {\bf 41} (2009) 1539
  [arXiv:0809.3850 [hep-th]].
\bibitem{Hajicek} P.~H\'a\'{\j}i\v{c}ek, 
Commun. Math. Phys. 34, 37 (1973)
\bibitem{Virbhadra:1997ie}
  K.~S.~Virbhadra,
  Int.\ J.\ Mod.\ Phys.\ A {\bf 12} (1997) 4831
  [gr-qc/9701021].
  \bibitem{Janis:1968zz}
  A.~I.~Janis, E.~T.~Newman and J.~Winicour,
  Phys.\ Rev.\ Lett.\  {\bf 20} (1968) 878.
  \bibitem{Wyman} M. Wyman 
  PRD {\bf24}, 4, 839 (1981)
  \bibitem{Buchdahl}H. A. Buchdahl, Phys. Rev. {\bf115}, 1325 (1959). 
\bibitem{BBM} Bocharova N., Bronnikov K. and Melnikov V., Vestn. Mosk. Univ. Fiz. Astron. {\bf 6}, 706 (1970)

\end{thebibliography}
\end{document}